\documentclass[twocolumn,floatfix,groupedaddress,superscriptaddress,prb,showpacs]{revtex4}

\usepackage{graphicx}
\usepackage{amsmath}
\usepackage{amssymb}
\usepackage{amsfonts}
\usepackage{dcolumn}
\usepackage{bbm}
\usepackage{float}

\usepackage{color}

\begin{document}
\title{Dynamical Transition in Interaction Quenches of the One-Dimensional Hubbard Model}

\author{Simone A. Hamerla}
\email{simone.hamerla@tu-dortmund.de}
\affiliation{Lehrstuhl f\"{u}r Theoretische Physik I, 
Technische Universit\"{a}t Dortmund,
 Otto-Hahn Stra\ss{}e 4, 44221 Dortmund, Germany}

\author{G\"otz S. Uhrig}
\email{goetz.uhrig@tu-dortmund.de}
\affiliation{Lehrstuhl f\"{u}r Theoretische Physik I, 
Technische Universit\"{a}t Dortmund,
 Otto-Hahn Stra\ss{}e 4, 44221 Dortmund, Germany}

\date{\textrm{\today}}

\begin{abstract} 
We show that the non-equilibrium time-evolution after interaction quenches 
in the one dimensional, integrable Hubbard model
exhibits a dynamical transition in the half-filled case.
This transition ceases to exist upon doping.
Our study is based on systematically extended equations of motion.
Thus it is controlled for small and moderate times;
no relaxation effects are neglected. Remarkable similarities to the
quench dynamics in the infinite dimensional Hubbard model
are found suggesting dynamical transitions to
be a general feature of quenches in such models.
\end{abstract}



\maketitle

\section{Introduction}

In the last years, seminal experimental setups have been developed combining a very good decoupling of the quantum systems from their environment with a high degree of controllability of the system's parameters. This renders the observation of the temporal evolution of closed quantum systems
for long times possible. Switching the internal parameters provides tools to 
investigate systems out of equilibrium.
In optical lattices, the internal parameters such as hopping and particle-particle interaction
can be manipulated \cite{greiner02,greiner02b,kinoshita06,bloch08}.
Furthermore, various pumb-probe experiments based on 
ultrafast spectroscopy \cite{perfetti06} have been developed \cite{bloch08}.

These developments have triggered extensive theoretical studies of physics far from equilibrium,
based on a large variety of analytical and numerical tools 
\cite{anders05,cazal06,freericks06,kolla07,manmana07,moe08,eck09,uhr09,enss12,goth12}.
The goal is to qualitatively understand and to quantitatively describe the
dynamics of quantum systems far from thermal equilibrium. How do such states evolve? 
Do they relax towards equilibrium? How does this happen and on which time scales?

These issues are relevant in all dimensions. But so far the
infinite dimensional and the one-dimensional (1D) case have attracted the greatest interest.
Mainly, this is due to the possibility of approximation-free results in both cases.
The infinite dimensional case ($d\to \infty$) is amenable by dynamical mean-field theory
(DMFT) \cite{metzn89a,mulle89a,georg96} for which it is exact, also in the
non-equilibrium case \cite{freericks06}. 
The 1D case is amenable to  analytical (e.g.\ bosonization \cite{cazal06,uhr09,dora11,rentr12}) 
and numerical  (e.g.\
 time-dependent density-matrix renormalization \cite{daley04,white04,kolla07,manmana07})  approaches.
In addition, there is a fascinating conceptual issue inherent to 1D systems:
All extended non-trivial \emph{integrable} quantum systems are 1D 
and the existence of an exhaustive number of conserved quantities influences the dynamics 
strongly because it implies a macroscopic number of conserved quantitities \cite{rigol07}.
Still, relaxation in local quantities due to dephasing may occur \cite{barth08}.

One efficient way to realize states far from equilibrium are interaction quenches \cite{barth08}. The
interaction value is changed abruptly increasing it \cite{cazal06,kolla07,manmana07,moe08,uhr09,dora11} 
or decreasing it \cite{goth12}.
We focus on interactions which are suddenly switched on with Fermi seas as initial states.
This simplifies the evaluation because all correlations are known analytically,
but it is not essential for the applicability of our approach which is based on extended
equations of motion \cite{uhr09}.

For an understanding of the dynamics of quenched systems 
also the behavior on short and moderate times directly after the quench matters. 
Relaxation, if it happens, may set in directly or coherent oscillations may take place
before relaxation occurs on a longer time scale. The occurrence of intermediate time scales, 
on which the system does not yet reflect the long-time behavior, is subsumed under the 
term ``prethermalization'' \cite{berge04,moe08,moeck09a}.
These issues are particularly difficult to study in strongly correlated systems. 

Interestingly, evaluating the DMFT equations after interaction quenches of 
the half-filled system two regimes
occurred \cite{eck09} which are characterized by qualitative differences in the time dependence
of the jump $\Delta n(t)$ in the momentum distribution at the Fermi level. For small values of the
interaction $U$, this quantity decreases gradually from its initial value of unity corresponding
to the non-interacting Fermi sea and a prethermalization plateau can
be discerned. For large values of $U$, the jump $\Delta n(t)$ is characterized by strong oscillations.
The key difference is that $\Delta n(t)$ displays zeros in the strong-interaction regime while
they are absent for weak interaction.

This observation of two qualitatively different dynamics could be retrieved in
a variational Gutzwiller approach, evaluated in Gutzwiller approximation, by Schir\'o
and Fabrizio \cite{schir10}. Their treatment maps the quantum dynamics
to a classical mechanics problem. True relaxation, however, is neglected in this way.
But the resulting equations allow for an analytic evaluation.
The oscillatory regime for weak quenches and the one for strong 
quenches are separated by a singularity indicating a dynamical transition.
Note that the Gutzwiller approximation becomes exact for infinite
coordination number \cite{metzn88,gebha90} so that the similarity
between the DMFT and the Gutzwiller results may not suprise.

In view of these results, it is our goal to show that a Hubbard model
with completely different properties displays essentially the
same dynamical transition. Thus we study the one-dimensional Hubbard model.
It is different from the Hubbard model examined by DMFT and Gutzwiller techniques
in two important aspects: (i) Scattering between the excitations is controlled
by momentum conservation in 1D \cite{luthe75,halda80,meden92,voit95,miran03}
while momentum conservation is suppressed at internal 
vertices in large dimensions \cite{metzn89a,mulle89a}.
(ii) The macroscopic number of conserved quantities in the integrable 1D
Hubbard model \cite{essle05} should constrain the non-equilibrium dynamics
so that it differs significantly from what is seen in higher dimensions.
Yet we find that the dynamical transition is also present 
in 1D with even quantitative similarities.

The article is set up as follows. Next, the model, the quench, and our approach
to them are introduced. In Sect.\ III, the results for the time-dependent jump
in the momentum distribution are presented. In Sect.\ IV, we analyse
the data with respect to the occurrence of a dynamical transition.
Finally, conclusions are drawn in Sect.\ V.

\section{Quenched Hubbard Model}

For the quench, we start from a Fermi sea and switch on the interaction abruptly. 
Thus the Hubbard Hamiltonian becomes time-dependent and reads
\begin{align}
\label{eq:Hu}
H_\text{Hu} = -J \sum_{\langle i,j;\sigma\rangle} 
\big(\hat{c}_{i,\sigma}^\dagger\hat{c}_{j,\sigma}^{\phantom\dagger} + \text{h.c.}\big)
+U(t)\sum_i \hat n_{i,\uparrow} \hat n_{i,\downarrow}
\end{align}
with a local repulsion $U(t)$,
where $\hat{c}_{j}^\dagger$ ($\hat{c}_{j}^{\phantom\dagger}$) create (annihilate) a particle
with spin $\sigma$ at site $j$ and $\hat n_j=\hat{c}_{j}^\dagger \hat{c}_{j}^{\phantom\dagger}$.
 We study $U(t)=\Theta(t)U\ge 0$ and define  the band width $W=4J$ as the natural energy scale.

The approach used is a systematically controlled expansion of the Heisenberg equations of motion
for $\hat{c}_{j,\sigma}^\dagger$  \cite{uhr09}. By commuting the interacting $H$
after the quench recursively with $\hat{c}_{j,\sigma}^\dagger$, i.e., by
applying the Liouvillian,  we obtain 
differential equations for the prefactors of the expansion of $\hat{c}_{j,\sigma}^\dagger(t)$
in more and more monomials of $\hat{c}_{i,\sigma}^\dagger$ and $\hat{c}_{j}^{\phantom\dagger}$
at $t=0$. We perform this calculation in real space. The application of one commutation is a `loop'.
Roughly, each loop multiplies the number of  tracked monomials 
by a factor 3. We are able to realize up to 11 loops. In Fig.\ \ref{fig:varU}
the convergence of time-dependent results with the loop number is shown.
Our approach bears similarities to recent calculations based on recursively
constructed Hilbert spaces \cite{bonca07}. But we stress that our approach is
operator-oriented. The evaluation of expectation values is done only at the end at the
time instant of interest.

The evaluation of the differential equations on the one-loop level 
describes the time dependence of the initial operator $\hat{c}_{j,\sigma}^\dagger$ 
exactly in linear order in $t$. 
By each loop we increment the depth of the hierarchy of the equations of motions
by one and thus the time dependence of the initial operator $\hat{c}_{j,\sigma}^\dagger$
is precisely captured up to order $t^{n}$ for $n$ loops. Finally, we solve
the differential equations numerically  so that also higher powers of $t$
are generated. The conceptual asset of the approach is that it works 
directly on the infinite lattice by exploiting translational invariance.
No relaxation effects occurring up to the considered order are neglected.
A quantitative analysis of the convergence of the approach and of its
accuracy is presented in App.\ \ref{sec_tech}.

\section{Time-dependent jump in the momentum distribution}

\begin{figure}[ht]
    \begin{center}
    \includegraphics[width=0.95\columnwidth,clip]{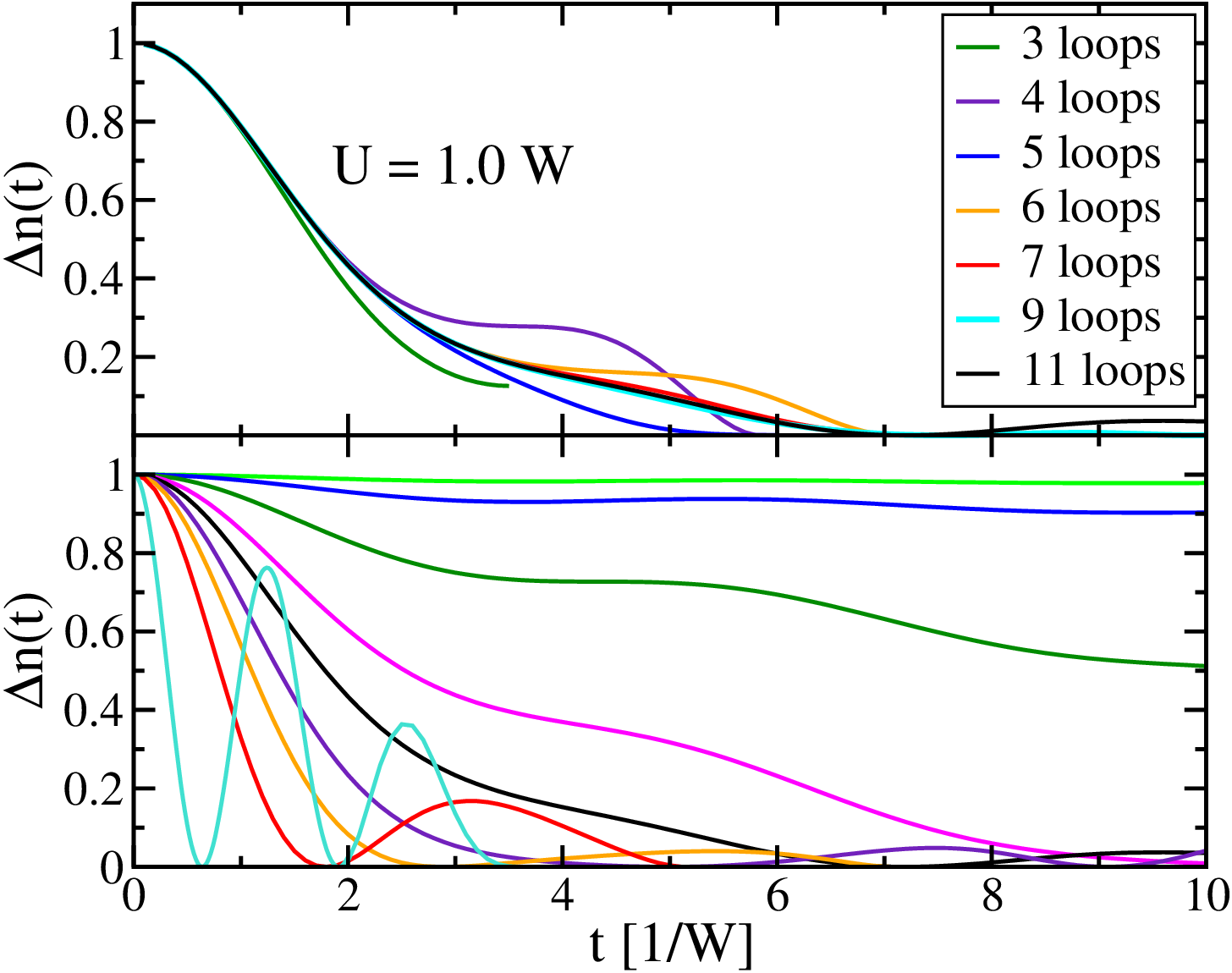}
    \end{center}
    \caption{(color online) Upper panel:
      Jump $\Delta n$(t) for the half-filled Hubbard model for various loop numbers. 
      Lower panel: 
      $\Delta n(t)$ for increasing $U$ (from top to bottom at small $t$: $U/W=0.125, 0.25, 0.5, 0.8, 1.0,
      1.25, 1.5, 2.0, 5.0$) in 11 loops.
      \label{fig:varU}
}
\end{figure}

We focus on the momentum distribution $n_k(t):=\langle \hat{c}_{k,\sigma}^\dagger
 \hat{c}_{k,\sigma}^{\phantom\dagger}\rangle(t)$ where $k$ is the wave vector 
 \cite{uhr09}. In particular,  we study
the jump $\Delta n(t):=n_{k_\text{F}+0}(t) - n_{k_\text{F}-0}(t)$ at the Fermi vector  $k_\text{F}$
with  $\Delta n(0)=1$ for the initial Fermi sea. After the quench, $\Delta n$ shows slow relaxation to 
zero and damped oscillations. The upper panel of Fig.\ \ref{fig:varU} 
depicts the jumps for increasing number of loops for the
half-filled Hubbard model. Good convergence is obtained for 11 loops up to 
about $t\approx 10/W$. The precise value up to which the data is reliable depends on the details, see
App.\ \ref{sec_tech}.
The lower panel of Fig.\ \ref{fig:varU} displays the data for various values of $U$.

The slow relaxation seen in Fig.\ \ref{fig:varU} is characteristic for a 1D model
as can be understood from results for Tomonaga-Luttinger models obtained
by bosonization \cite{cazal06,uhr09,dora11,rentr12}. It is found that
power laws instead of exponential relaxation occur. Our data agrees with this
expectation see data in Ref.\ \onlinecite{hamer12}.
 
\begin{figure}[ht]
    \begin{center}
    \includegraphics[width=0.95\columnwidth,clip]{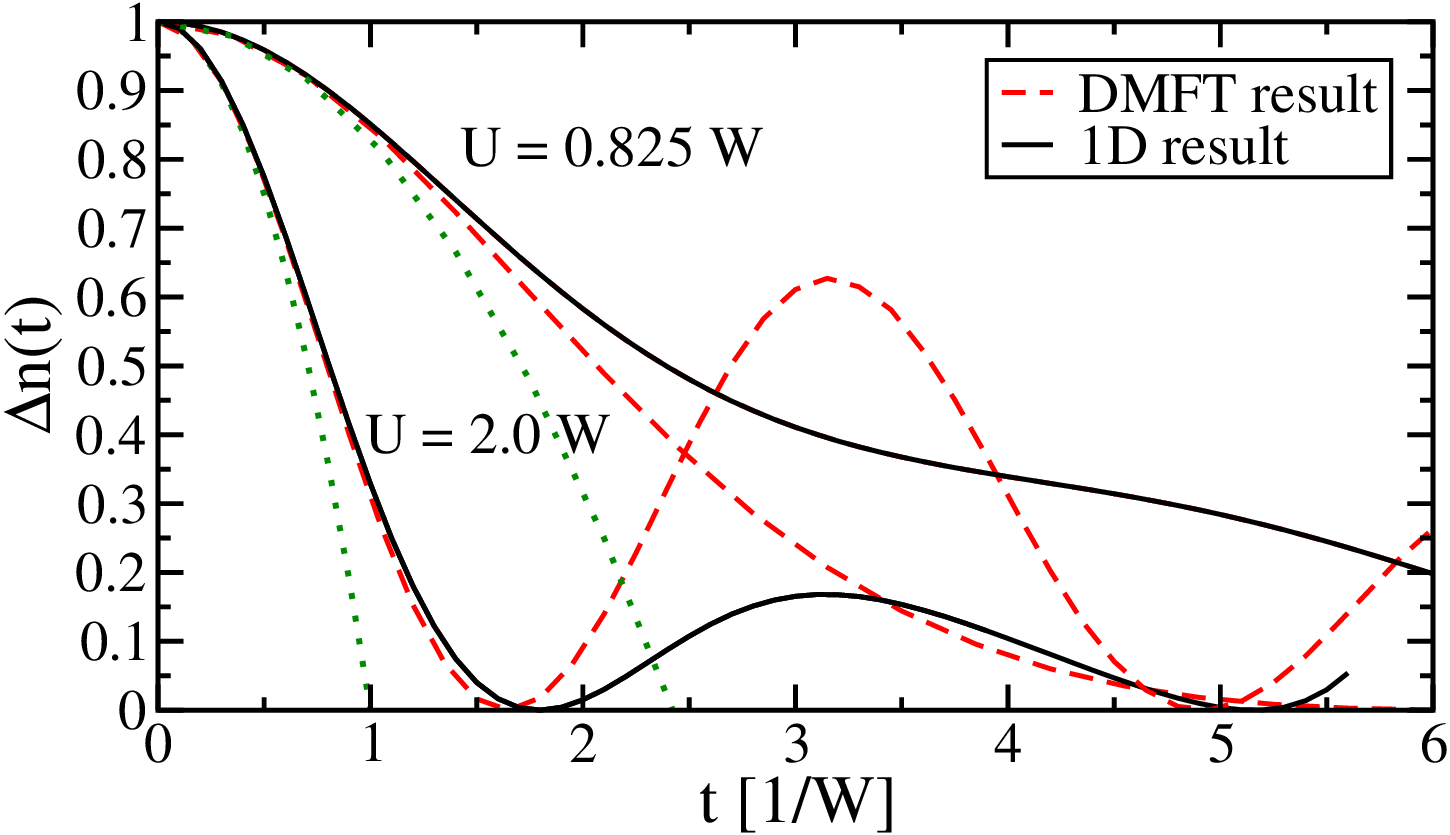}
    \end{center}
    \caption{(color online) 
       Jump $\Delta n(t)$ for the half-filled Hubbard model in 1D (black, solid lines)
       and on the Bethe lattice with infinite coordination number 
       (gray, dashed lines from Ref.\ \cite{eck09}).
       Dotted line: Second order result $\Delta n = 1- U^2t^2/4$.
      \label{fig:dmft}
}
\end{figure}

Our first key observation stems from the comparison of the 1D data
with exact DMFT data for a Bethe lattice \cite{eck09}. 
The common lore expects
a crucially distinct  behavior due to the differing dimensionality and
the integrability of the 1D Hubbard model \cite{essle05}. The scattering in 1D is 
strongly restricted due to momentum conservation 
\cite{luthe75,halda80,meden92,voit95,miran03} 
while this conservation is irrelevant at internal vertices
in DMFT \cite{metzn89a,mulle89a}. Yet Fig.\
\ref{fig:dmft} shows qualitatively similar results for larger values of $U$. 
For  $U\lessapprox W$ quantitative differences prevail beyond
$t=2/W$ so that it is difficult to discern qualitative similarities.
But for interaction values beyond the band width $W$ coherent oscillations in $\Delta n(t)$ 
occur \cite{eck09} which agree in that the minima appear at almost
the same values.  To stress that this is not simply an effect of the leading 
order in $t$  the second order result $\Delta n(t)=1- U^2 n(2-n) t^2/4$
($n$ filling factor) is included in Fig.\ \ref{fig:dmft}. 
The amplitudes of the oscillations, however, differ significantly.
We conclude that the periods of the oscillations
are a candidate for similarities while the amplitudes and 
their decay are more strongly depending on the model. 
In 1D, a slow power law relaxation is generic
while exponential relaxation is expected in higher dimensions in
agreement with the DMFT data. These striking similarities indicate that for large $U$
the physics is governed by local processes as is explained below.

\section{Dynamical transition}

We return to the oscillations $\propto \cos(2\pi t/T)$ 
which appear both for strong quenches and for weak
quenches. Schir\'o and Fabrizio derived an analytic formula for the period $T$
of the oscillations at half-filling
\begin{equation}
T=\left\{
\begin{array}{lcl}
\frac{4\sqrt{2}}{U_c}K(2U/U_c) & \text{for} & U<U_c
\\
\frac{4}{U_c}K(U_c/2U) & \text{for} & U>U_c,
\end{array}
\right.
\label{eq:schiro}
\end{equation}
where $K(x)$ is the complete elliptic integral of the first kind and $U_c$ the
value where the Mott transition occurs in the Gutzwiller approach  \cite{schir10}.
It is given by $-8 E_\text{kin}$ where $E_\text{kin}$ is the kinetic energy
of the half-filled Fermi sea. 
Away from half-filling, the logarithmic singularity
in \eqref{eq:schiro} is smeared out.

\begin{figure}[ht]
    \begin{center}
    \includegraphics[width=0.95\columnwidth,clip]{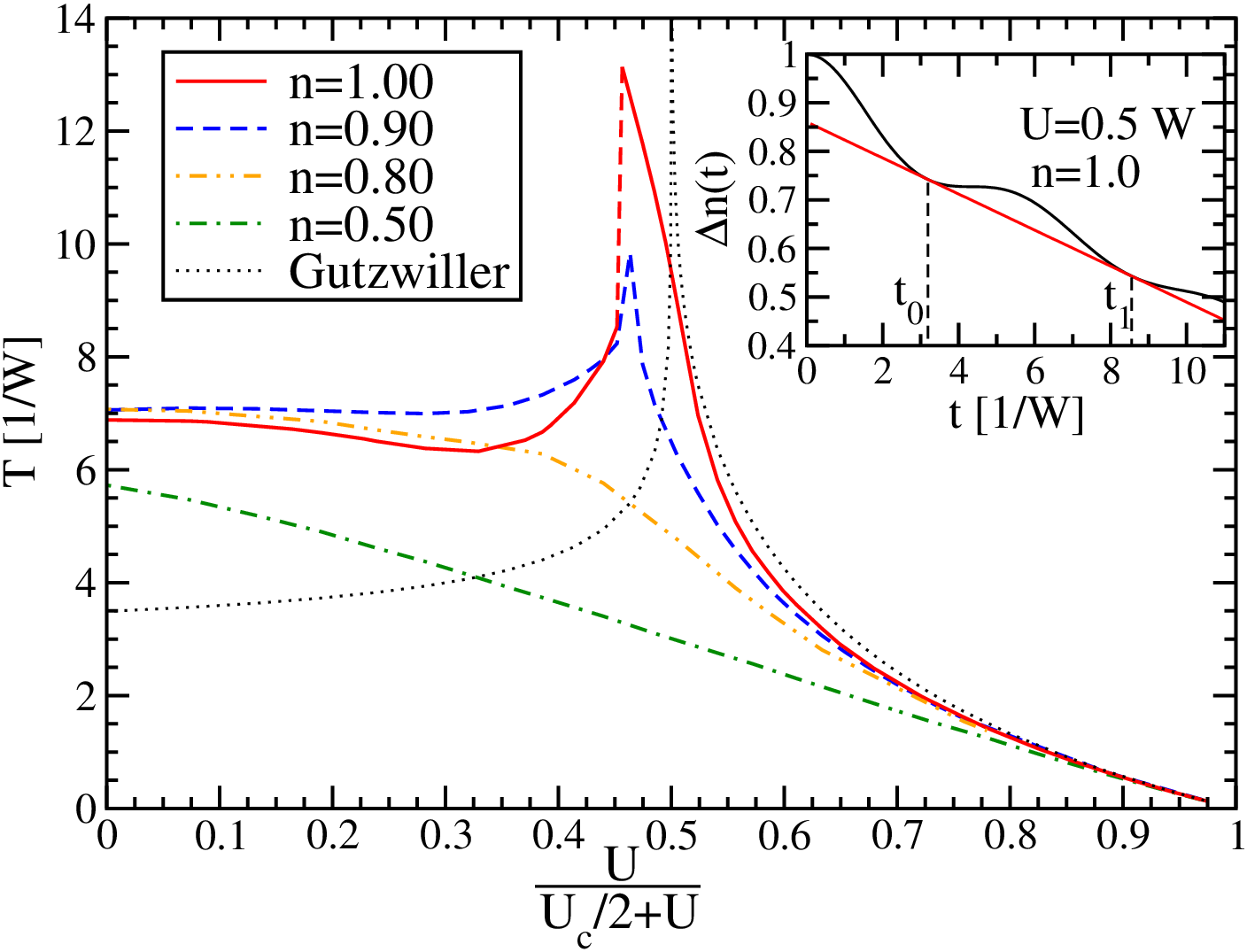}
    \end{center}
    \caption{(color online) 
      Period of the oscillations in $\Delta n(t)$ 
      as function of $U$ ($U_c=8W/\pi$ in 1D); $n$ denotes the total filling factor per site.
      The dotted curve depicts the semiclassical Gutzwiller result \eqref{eq:schiro} 
      \cite{schir10}.
      \label{fig:period}
}
\end{figure}

What does our data yield for the oscillation periods?
In absence of relaxation, thus without changes in the oscillatory amplitudes, it is
straightforward to read off the period $T$. In our data, this is less obvious.
To determine the period $T$ we fit a straight line to the data such
that it is a double tangent to two points close to the first two minima, see inset
of Fig.\ \ref{fig:period}. From the first instant we deduce the period
 $T=2t_0$. The double tangent is constructed to take into account that the
 oscillatory part may be small and sit on top of a non-oscillatory relaxation.
 In case only one minimum can be determined, we read off its instant in time to find $t_0$.

Representative results are depicted in Fig.\ \ref{fig:period} for various
fillings. For comparison, the analytic formula \eqref{eq:schiro} is plotted as well.
 At and around half-filling ($n=1$) and for interaction
values $U$ smaller than about $U_c/2$ the oscillation period $T$ hardly depends
on $U$. The rather flat curve in this regime
 is in agreement with the semiclassical finding for half-filling \eqref{eq:schiro}
 although the value of $T$ differs roughly by a factor of two which may not
 surprise in view of the very different models. Note that $T$ stays finite for $U\to 0$.
Its value in this limit depends on the filling. Data for other fillings can be found
in App.\ \ref{sec_osc} and in Figs.\ \ref{fig:doping} and \ref{fig:period} below. 
We stress that all oscillations in the weak quench
regime do no show zeros in $\Delta n(t)$ so that $\Delta n(t)>0$.

For quenches to large $U$ and in particular for $U\to\infty$ the curves for all
fillings converge well to the asymptotic behavior $T=2\pi/U$. This value corresponds to 
Rabi oscillations in the local two-level system which is given 
by a local singly and doubly occupied site. In other words, the lattice
behaves as if it were made up from independent Hubbard atoms in first approximation.
This finding stems from the basic two-loop calculation leading to
\begin{align}
\nonumber
\hat{c}_{0,\uparrow}^\dagger(t)&=h_0(-1,t) :\hat{c}_{-1,\uparrow}^\dagger :+h_0(0,t) :\hat{c}_{0,\uparrow}^\dagger:
\\& +h_0(1,t) :\hat{c}_{1,\uparrow}^\dagger: +h_1(0,0,0,t)  :\hat{c}_{0,\uparrow}^\dagger\hat{c}_{0,\downarrow}^\dagger\hat{c}_{0,\downarrow}^{\phantom\dagger}:
\label{eq:2loop}
\end{align} 
where site 0 can be any site on the chain because of translational 
invariance. The coefficients follow the differential equations 
\begin{subequations}
\begin{align}
\nonumber
\partial_t h_0(0,t) &= -Jih_0(-1,t)-Jih_0(1,t)
\\& +U\frac{n}{2}\left(1-\frac{n}{2}\right)ih_1(0,0,0,t)\\
\partial_t h_0(-1,t) &= -Jih_0(0,t)\\
\partial_t h_0(1,t) &= -Jih_0(0,t)\\
\partial_t h_1(0,0,0,t) &= Uih_0(0,t) +(1-n)h_1(0,0,0,t)
\end{align}
\end{subequations}
for the coefficients. This result covers the leading order in $t$ independent of the value of 
$U$ because the commutation of the interaction part in \eqref{eq:Hu} with the local
operators, i.e, those at site 0, in \eqref{eq:2loop} does not yield any other operator term
than those that we included. In this sense, the approach used becomes exact for $U\to\infty$.
The surrounding lattice sites act as damping bath.

At half-filling, the above differential equations can be solved analytically and
we find for the oscillation period $T=2\pi/\sqrt{U^2+W^2/2}$. For finite doping
the numerical solution is easily done. In both cases, $T=2\pi/U$ is the leading order
result in an expansion in $1/U$. 
The equations of motion approach is well-controlled for large values of $U$
because the number of operators generated by commutation  with the interaction part
does not grow infinitely. Their number is finite for a fixed number of
sites involved because the corresponding Hilbert space is finite.
The subleading order $W/U^2$ is exactly captured if all operators acting on
two sites are included which happens in the seven-loop calculation.
Clearly, local processes dominate for large values of $U$.

Returning to the two regimes of quenches, 
we stress that only the oscillations at half-filling display zeros
 in $\Delta n(t)$ after the \emph{strong} quenches. For any finite doping 
 ($n\neq 1$) the oscillations persist, but do not reach zero any more in their
 minima as shown in detail in Fig.\ \ref{fig:doping}.
 Even 2\% of doping are sufficient to shift the first minimum from zero upwards to
a finite value within numerical accuracy, see right panel in Fig.\ \ref{fig:doping}.

\begin{figure}[ht]
    \begin{center}
    \includegraphics[width=0.95\columnwidth,clip]{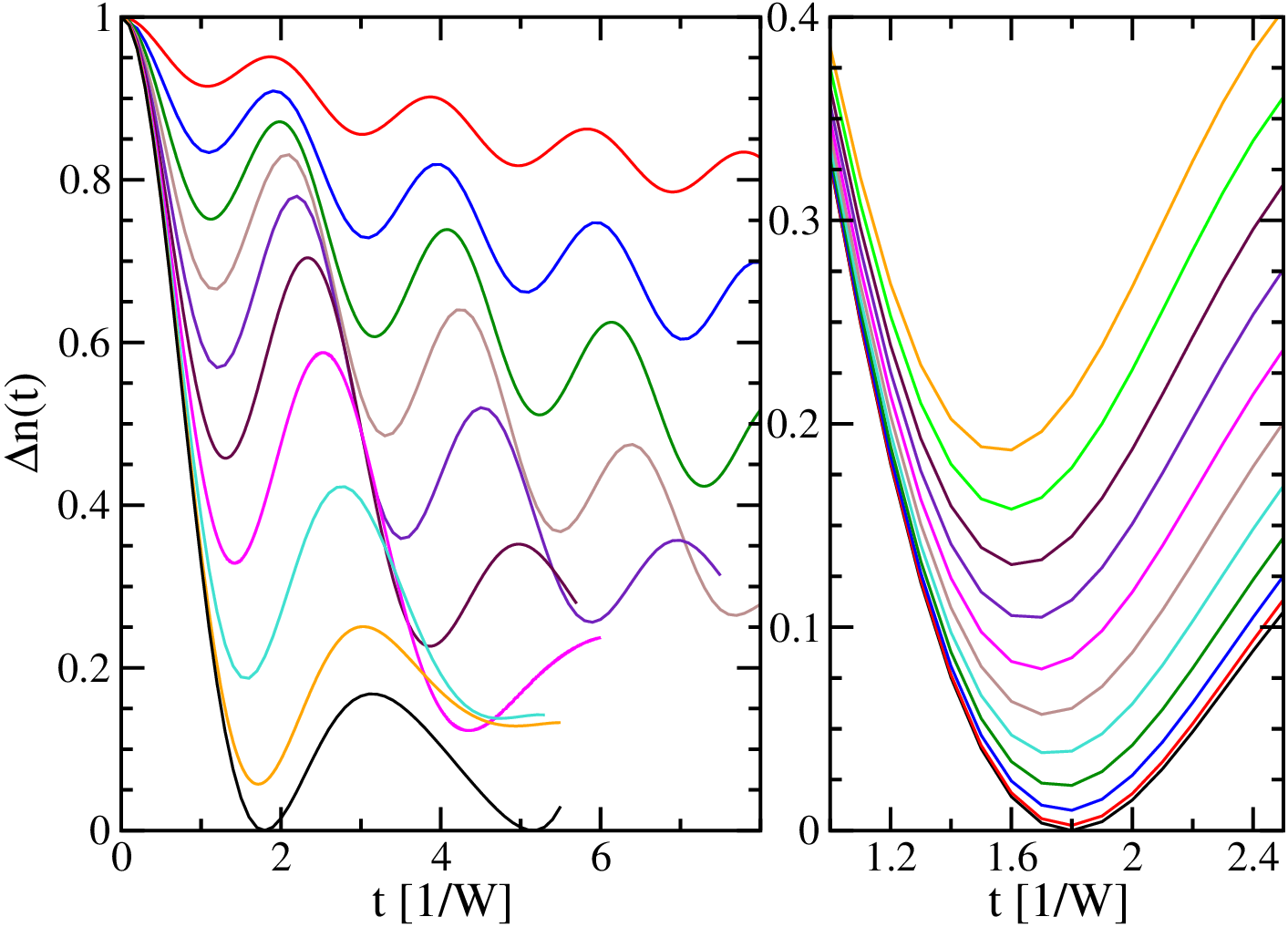}
    \end{center}
    \caption{(color online) 
  Oscillations in $\Delta n(t)$ for various dopings. In the left panel, 
  the dopings are from top to bottom at the first minimum:
  $n=0.1, 0.2, 0.3, 0.4, 0.5, 0.6, 0.7, 0.8, 0.9,  1.0$. 
  In the right panel, 
  the dopings are from top to bottom at the minimum:
  $n=0.8, 0.82, 0.84, 0.86, 0.88, 0.9, 0.92, 0.94, 0.96, 0.98,  1.0$. 
  Clearly, zeros occur only at half-filling.
      \label{fig:doping}
}
\end{figure}

The regimes of weak and of strong quenches are separated by
an anomaly catching the eye in Fig.\ \ref{fig:period} at
$U\approx U_c/2$. 
From our data we cannot identify a singularity
because there are some uncertainties in the determination of $T$, see
App.\ \ref{sec_err} for error bars and further discussion, and 
we cannot make statement about times beyond $\approx 15/W$.
Nevertheless, our finding is suggestive of a singular behavior which
may be a jump (indicated by a dashed line) 
or a logarithmic singularity (the periods rise by about a factor 
two above their $U\to0$ value). The similarity to
the semiclassical result is surprising and indicates that indeed
the two regimes of quenches are separated by a dynamical transition. 
The small shift of the position of 
the anomaly from $U=U_c/2$ to $U\approx 0.43 U_c$ is of minor importance.

We emphasize that all oscillations after strong quenches, to the 
right of the dashed part of the half-filled curve, display zeros   $\Delta n(t)$ while no
zeros are found for weak quenches, to the left of the dashed part. This holds, however,
only at half-filling. For finite doping, even strong quenches do not imply zeros
so that the transition ceases to exist for any doping, see Fig.\ \ref{fig:doping}.
Thus the data constitutes very strong evidence for a qualitative, dynamical transition
between both regimes at half-filling. At finite doping, it appears to be
washed out immediately. The anomaly, however, disappears gradually: We see 
in Fig.\ \ref{fig:period} that about 20\% of doping are required to make the anomaly vanish completely,
cf.\ App.\ \ref{sec_osc}.

The finding agrees with the semiclassical, relaxation-free Gutzwiller result 
in infinite dimensions
\cite{schir10} and with the available DMFT data \cite{eck09}.
Since such similar behavior occurs in extreme cases such as a one-dimensional
integrable model with important momentum conservation 
and an infinite dimensional model with suppressed momentum conservation 
the conclusion suggests itself that the dynamical transition
is a general feature in quenched, half-filled Hubbard models.

\section{Conclusions}

Summarizing, we studied interaction quenches in the one-dimensional Hubbard model
by an approach based on equations of motion. It is systematically controlled
in the depth of the hierarchy; here up to 11 commutations were carried out.
The accuracy of the data is systematically controlled by the  comparison of
the results for different loop numbers.

We analyse the time-dependence of the jump in the momentum distribution
starting from the initial Fermi sea. Slowly decaying
oscillations are found. At half-filling, two qualitatively different regimes
appear: For strong quenches, the jumps display zeros, for weak quenches they do not.
In the dependence of the oscillation period $T$ on the interaction  $U$ 
a clear anomaly appears separating both regimes. It indicates a dynamical transition.
Since this scenario is in surprising agreement with previous findings in infinite-dimensional
models \cite{eck09,schir10} we suggest that this scenario holds
for half-filled Hubbbard models in general.

\begin{acknowledgments}
We are indebted to M.\ Eckstein for providing the DMFT data and to S.\ Kehrein, M.\ Kollar, 
V.\  Meden, M.\ Moeckel, M.\ Schir\'o, and J. Stolze for useful discussions.
We acknowledge support by the Studienstiftung des deutschen Volkes (SAH)
and by the Mercator Stiftung (GSU).
\end{acknowledgments}


\appendix

\section{Technicalities}
\label{sec_tech}

We are interested in momentum distributions after interaction quenches.
They can be computed by Fourier transformation of the one-particle equal time propagators
\begin{align}
G(\vec{r},t) = \langle \text{FS}|\hat{c}(\vec{r},t)^{\phantom\dagger}\hat{c}^\dagger(0,t)|\text{FS}\rangle
\end{align}
where the expectation value is taken with respect to the noninteracting Fermi sea $|\text{FS}\rangle$
which represents the initial state before the quench. The $\vec{r}$ stands for
a site on the lattice under study, here a chain in one dimension (1D).
The time-dependent operators $\hat{c}$ and $\hat{c}^\dagger$ are represented by the following ansatz
\begin{align}
\hat{c}^\dagger(\vec{r},t) = \hat{T}_{\vec{r}}^\dagger + 
\left(\hat{T}^\dagger \hat{T}^\dagger\hat{L}^\dagger\right)_{\vec{r}} + ...
\label{ansatz}
\end{align}
where $\hat{T}^\dagger$ ($\hat{L}^\dagger$) denote various superpositions of 
particle (hole) creation operators. For instance, the single particle creation $\hat{T}^\dagger$ 
is given by
\begin{align}
\hat{T}_{\vec{r}}^\dagger = \sum_{|\vec{\delta}| \lessapprox v_{\text{max}} t} \sum_\sigma h_0(\vec{\delta}, t) \hat{c}_{\vec{r}+\vec{\delta},\sigma}^\dagger
\end{align} 
with time-dependent prefactors $h_0(\vec{\delta},t)$.
The shifts $\vec{\delta}$ can be bounded from above by  $v_{\text{max}} t$ 
where $v_{\text{max}}$ is the maximal velocity in the sense of the Lieb-Robinson theorem \cite{lieb72,cala06}.
Physically, this means that there is a maximum velocity with which the essential effects of
the Hamiltonian after the quench can travel. Of course, exponentially small effects
will be neglected. Hence, in order to describe the dynamics correctly up a certain time one has
to include processes up to a certain spatial range \cite{uhr09}. This idea is also used in other approaches
\cite{enss12}.

To calculate the time dependence of the prefactors we use the Heisenberg equation 
$\partial_t \hat{{A}}(\vec{r},t) = i\left[\hat{{H}},\hat{{A}}(\vec{r},t)\right]$ for the time derivative of an operator $\hat{A}$. On calculating the commutator 
$\left[\hat{{H}},\hat{c}^\dagger(\vec{r},t)\right]$ 
we encounter two cases. The commutation of the noninteracting part 
of the Hamiltonian $\hat{{H}}_0$ leads to a shift of the fermionic operators
in space, whereas the commutation with the interaction term $\hat{{H}}_{\text{int}}$
may additionally create or annihilate particle-hole pairs 
$\hat{T}^\dagger \hat{L}^\dagger$. 
Iterating this process then leads to  the ansatz \eqref{ansatz}.
With each commutation more terms with higher number of particles involved are created. Thus the amount of terms grows exponentially: At 11 loops we deal with up to $5\cdot 10^5$
monomials in the Hubbard model and a set of differential equations
with about $2\cdot 10^7 $  terms on the right hand side.

The differential equations of the prefactors can be solved numerically with the initial conditions
 $h_0(0,0) =1$ and $h_i(\vec{r},t)=0\; \forall i\neq 0$�.
Because each commutation comprises one order in time 
$t$ a calculation with $n$ commutations provides results for $\hat{c}^\dagger(t)$ which are exact up to order $t^n$. We stress, however, that we do not use a series expansion in $t$ but
solve the (truncated) set of differential equations numerically.

Due to the proliferating number of additional terms arising within the calculation we have to restrict ourselves to a finite number of commutations.
As terms appearing during the last commutation for the first time lead to an overestimation of the weight loss for the one particle terms, we omit them to improve the convergence. 
A calculation with $m$ commutations performed in this way is called an $m$-loop calculation.

\begin{figure}[ht]
    \begin{center}
    \includegraphics[width=0.95\columnwidth,clip]{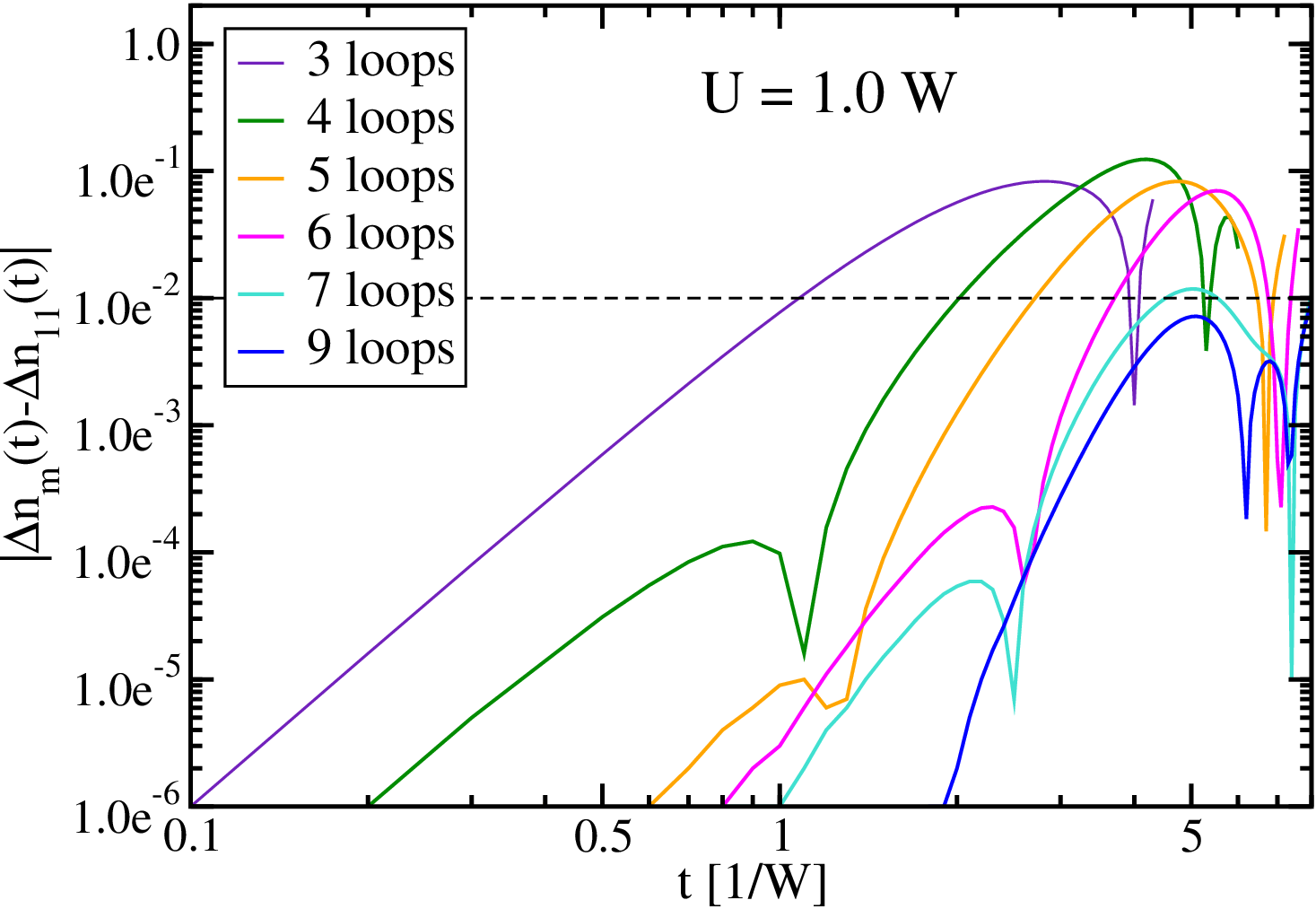}
    \end{center}
    \caption{(color online) 
  Absolute difference of the jump $\Delta n_m(t)$ at various numbers of loops $m$ 
  relative to the $11$-loop result $\Delta n_{11}(t)$   
  for the half-filled Hubbard model. 
  Dashed line: Threshold for the determination of the runaway time.
      \label{fig:supp_conv1}
}
\end{figure}

To illustrate how each commutation improves the result the absolute difference between an $m$-loop calculation and the $11$-loop calculation is shown for a quench with $U=1.0W$ in a double logarithmic plot in Fig.\ \ref{fig:supp_conv1}.
Based on this difference a runaway time is defined as the time $t$ beyond which the difference takes values larger than a certain threshold which we set to $0.01$. 
This threshold is depicted in Fig.\ \ref{fig:supp_conv1} as dashed line.

\begin{figure}[ht]
    \begin{center}
    \includegraphics[width=0.95\columnwidth,clip]{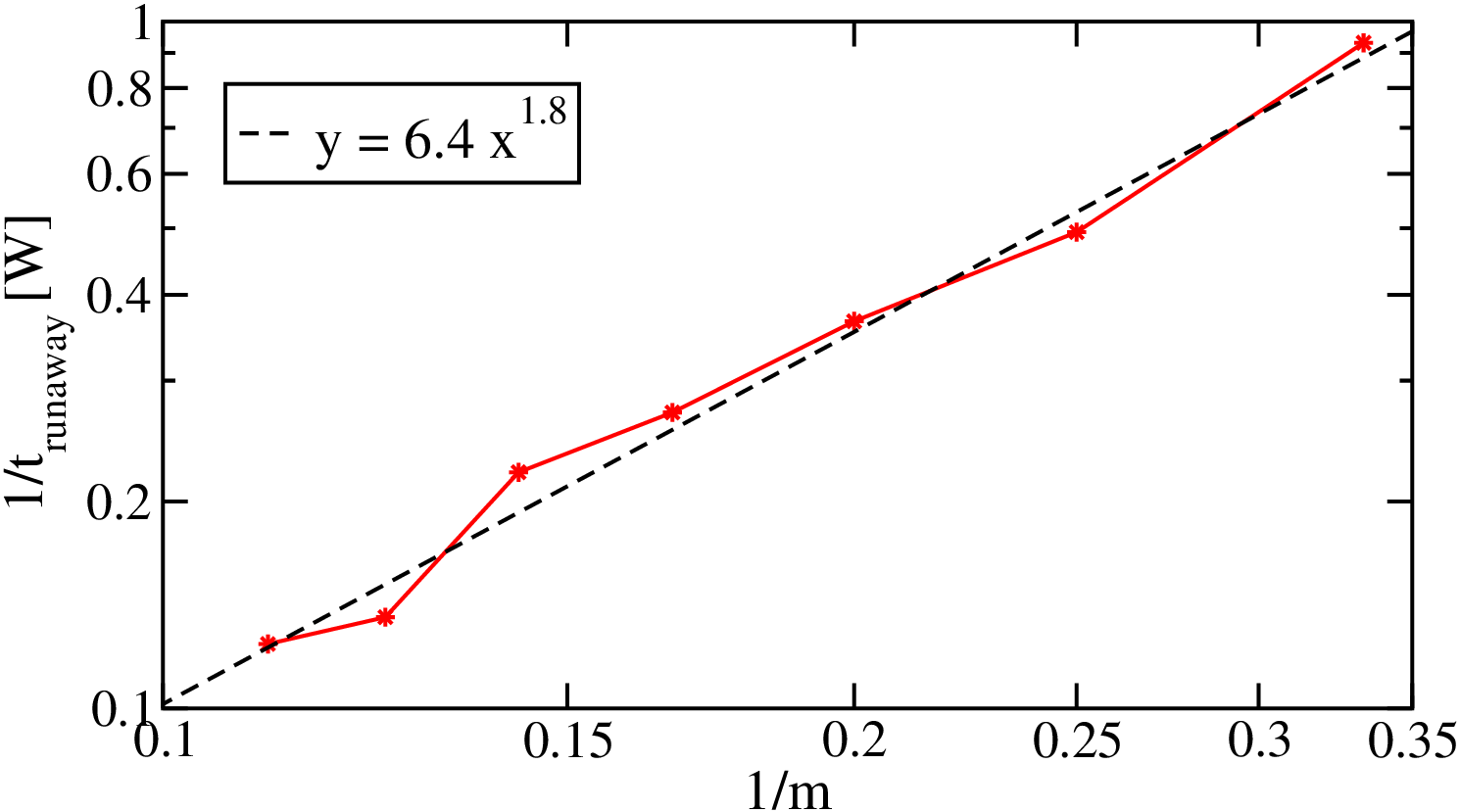}
    \end{center}
    \caption{(color online) 
   Double logarithmic plot of the inverse runaway time vs.\ the inverse number of loops of the corresponding calculation. Dashed line: Power law fit of the data with an exponent of about $1.8$.
      \label{fig:supp_conv2}
}
\end{figure}

The resulting inverse runaway time is shown in Fig.\ \ref{fig:supp_conv2} as function of $\frac{1}{m}$ where 
$m$ denotes the number of loops performed as described above. 
The curve shows a power law increase of the  runaway time as function of $\frac{1}{m}$,
i.e., a power law decrease of the inverse runaway time, with an exponent of about $1.8$. 
Note that this indicates that the converence is superlinear which is favorable for its
application in practical calculations.

\section{Oscillations for Various Filling Factors}
\label{sec_osc}

\begin{figure}[ht]
    \begin{center}
    \includegraphics[width=0.95\columnwidth,clip]{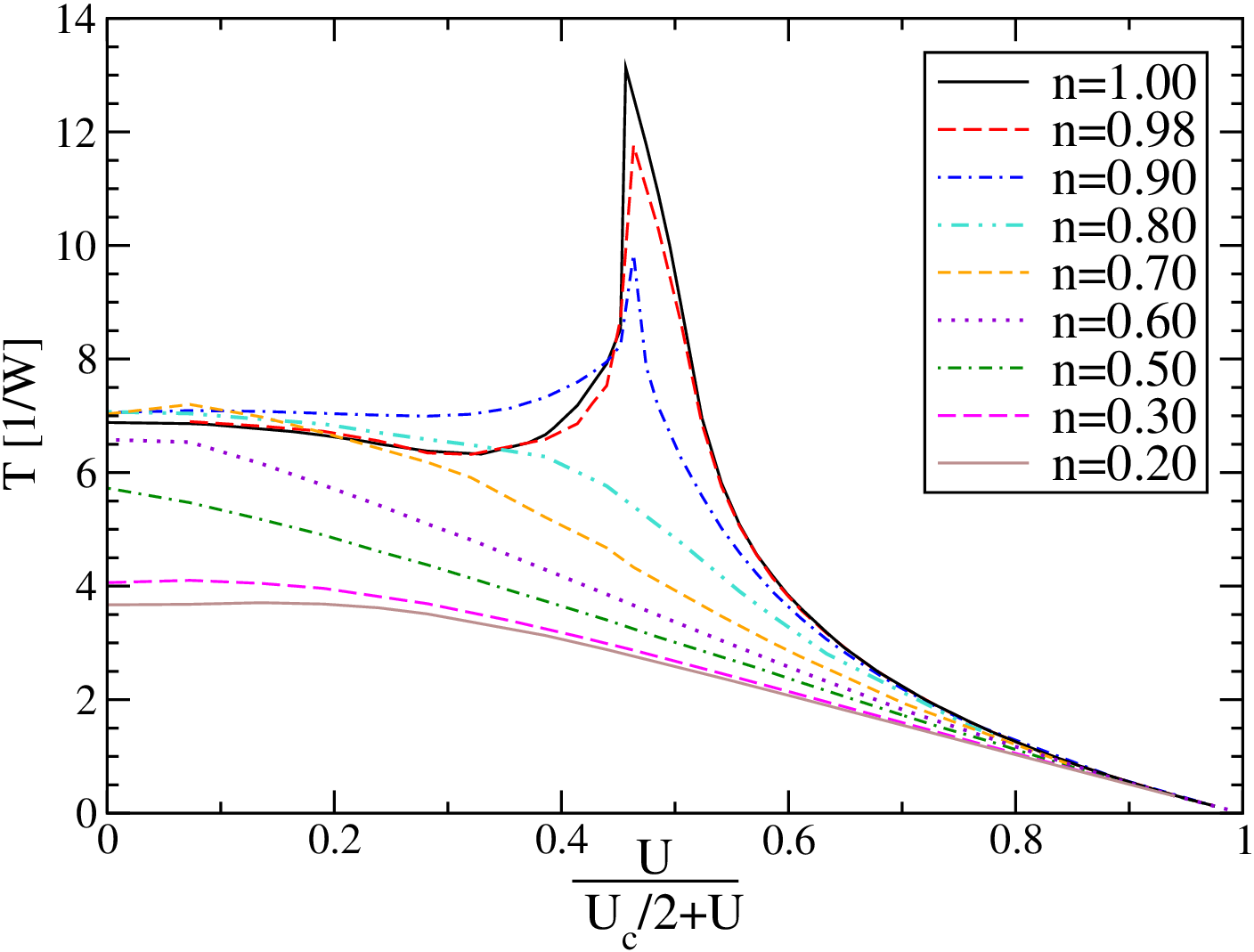}
    \end{center}
    \caption{(color online) 
  Period of the oscillations in $\Delta n(t)$ 
      as function of $U$ ($U_c=8W/\pi$ in 1D); $n$ denotes the total filling factor per site.
      \label{fig:supp_all}
}
\end{figure}

The period of oscillations for various values of the filling is shown in Fig.\ 
\ref{fig:supp_all} as function of the compactified interaction $\frac{U}{(U_c/2)+U}$.
The interaction $U_c$ denotes the values of the Hubbard interaction, where the Mott transition occurs
in the Gutzwiller approximation \cite{brink70,metzn88,gebha90,schir10}. 
It takes the value $8W/\pi$ in 1D.

Upon doping the anomaly at $U\approx U_c/2$ is gradually reduced until it vanishes completely at around
$n\approx 0.8$,  leaving behind a monotonic behavior of the period.
In contrast to the gradual disappearance of the anomaly, the zeros for strong quenches to the
right side of the anomaly vanish immediately upon doping. This was illustrated in Fig.\ \ref{fig:doping}.\\

\section{Systematic Errors in the Determination of the Period }
\label{sec_err}
In the determination of the periods of oscillation systematic 
inaccuracies due to the finite amount of loops occur.
There are two main sources: (i) The finite number of loops and (ii) the way the period $T$ is extracted
from the data. The effect of the finite number of loops is estimated 
by comparing results from the calculation with 11 loops
with calculations with a smaller number of loops. This yields estimate (i) for the systematic error.

In order to estimate the effect of the fitting procedure we compare the periods $T$ determined
from the double tangents described in the main article to an alternative analysis.
In this alternative procedure we read off the positions $t_0'$ of the first 
minimum of $\Delta n(t)$ and deduce $T'= 2t_0'$. The absolute value $|T-T'|$ yields
estimate (ii) for the systematic error.

\begin{figure}[ht]
    \begin{center}
    \includegraphics[width=0.95\columnwidth,clip]{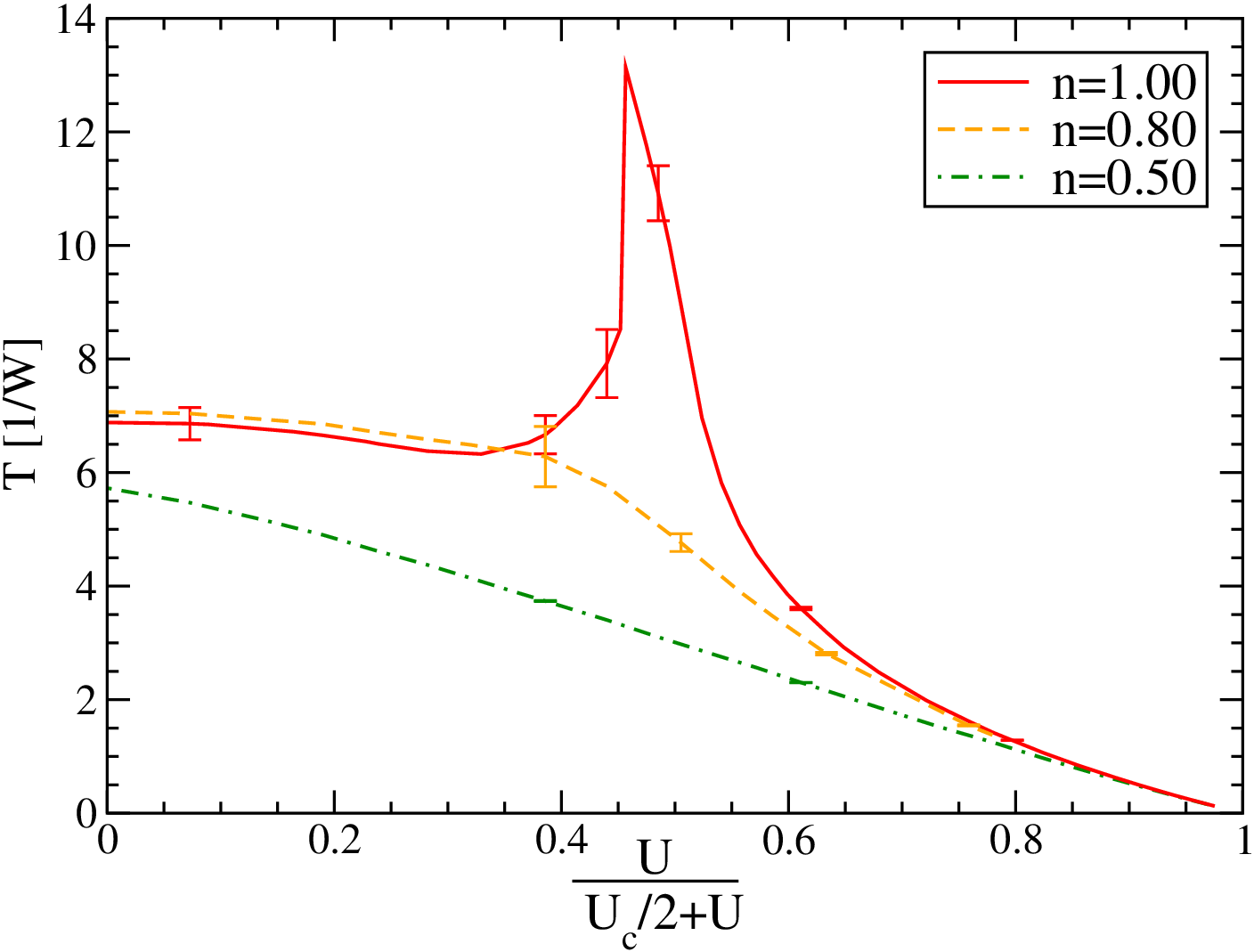}
    \end{center}
    \caption{(color online) 
  Period of the oscillations for the fillings $n=1.0, 0.8, 0.5$. The error bars depict the error of the calculated period for some exemplary values of the interaction. They are estimated as described in the
  main text.
      \label{fig:supp_error}
}
\end{figure}

Finally, the maximum of both estimates (i) and (ii) is used as the error of the period. 
These errors are shown by error bars for fillings $n=1.0, 0.8$ and $0.5$ for some exemplary values 
of the interaction in Fig.\ \ref{fig:supp_error}. In the regime of large interactions, the
errors are negligible. Around the anomaly, they are largest as could be expected.
But the position and shape of the anomaly are hardly affected by the systematic errors.
Thus we are confident that our conclusions are valid.

If a double tangent can be constructed as in the inset of Fig.\ 3 in the main article,
a third way to determine the period is by $T''=t_1-t_0$. This procedure is not possible
for all points so that we do not use it for the determination of the curves $T(U)$.
For small values of $U$, the third approach yields periods shorter by up to $1/W$.


\end{document}